\documentclass[prb]{revtex4-1}
\usepackage{graphicx}
\usepackage{dcolumn}
\usepackage{bm}
\usepackage{epstopdf}
\usepackage{color}
\usepackage{ulem}

\begin{document}

\title{Anderson-Bogoliubov and Carlson-Goldman modes in counterflow superconductors: Case
study of a double monolayer graphene.}

\author{K.\,V.\,Germash$^1$, D.\,V.\,Fil$^{1,2}$}

\email{fil@isc.kharkov.ua}

\affiliation{$^1$Institute for Single Crystals, National Academy of Sciences of Ukraine,
60 Nauky Avenue, Kharkiv 61072, Ukraine\\
$^2$V.N. Karazin Kharkiv National University, 4 Svobody Square, Kharkiv 61022, Ukraine}

\begin{abstract}
The impact of electron-hole pairing on the spectrum of plasma excitations in double layer
systems is investigated. The theory is developed with reference to a double monolayer
graphene. Taking into account the coupling of scalar potential oscillations with
oscillations of the order parameter $\Delta$, we show that the spectrum of antisymmetric
(acoustic) plasma excitations contains two modes: a weakly damped mode below the gap
$2\Delta$ and a strongly damped mode above the gap. The lower mode can be interpreted as
an analog of the Carlson-Goldman mode. This mode has an acoustic dispersion relation at
small wave vectors and it saturates at the level $2\Delta$ at large wave vectors.   Its
velocity is larger than the velocity of the Anderson-Bogoliubov mode
$v_{AB}=v_F$/$\sqrt{2}$, and it can be smaller than the Fermi velocity $v_F$. The damping
rate of this mode strongly increases under increase of temperature. Out-of-phase
oscillations of two order parameters in two spin subsystems are also considered. This
part of the spectrum contains two more modes. One of them is interpreted as an analog of
the Anderson-Bogoliubov (phase) mode and the other, as an analog of the Schmid
(amplitude) mode. With minor modifications the theory can be extended to describe
collective modes in a double bilayer graphene as well.
\end{abstract}

\maketitle

\section{Introduction}

Electron-hole pairing is a phenomenon analogous to the Cooper pairing that may occur in
double layer systems consisting of an electron-doped layer and a hole-doped
layer\cite{1,2} (see also Ref. \onlinecite{3} for a review). In the paired state the
system may support dissipationless counterflow - a flow of oppositely directed
superconducting electric currents in adjacent layers. The phenomenon is referred to as
the superfluidity of spatially indirect excitons, exciton condensation in bilayers, or
the counterflow superconductivity.

A strong increase of the counterflow conductivity at low temperature caused by the
electron-hole pairing was observed \cite{4,5,6} in quantum Hall bilayers with the total
filling factor of 1 ($\nu_T=2\pi\ell_B^2(n_1+n_2)=1$, where $n_i$ is the electron density
in the $i$-th layer and $\ell_B$ is the magnetic length). The current state of art in
experimental investigations of exciton condensation in $\nu_T=1$ quantum Hall bilayers is
described in Ref. \onlinecite{7}. Quantum Hall bilayers demonstrate a zero bias peak in
the differential tunneling conductance\cite{8} and a strong interlayer drag
resistance\cite{9}. These two features are  considered as experimental signatures of the
electron-hole pairing. Similar features were observed in  double layer systems in zero
magnetic field. The increase of the interlayer drag resistance at low temperature was
detected in a double quantum well in AlGaAs heterostructures \cite{10,11} and in hybrid
double layer systems comprising a monolayer (bilayer) graphene in close proximity to a
quantum well created in GaAs \cite{12}. Experimental observation of strongly enhanced
tunneling between two graphene bilayers  at  equal occupation of adjacent bilayers by
electrons and holes was reported recently\cite{13}. The registered tunneling conductance
at small bias voltage was many orders of magnitude greater than that predicted for
uncorrelated electrons and holes.

Theoretical consideration shows that promising candidates for a realization of
electron-hole pairing in zero magnetic field are double
monolayer\cite{14,15,16,17,18,19}, double bilayer\cite{20,21,22} and double
multilayer\cite{23} graphenes, double transition metal dichalcogenide
monolayers\cite{24,25,26}, a phosphorene double layer\cite{27,27-1} and topological
insulators\cite{28,29}.

In recent papers \cite{30,31,32} we have predicted the effects that can be considered as
additional hallmarks of the electron-hole pairing. It was shown \cite{30} that the
electron-hole pairing suppresses the ability of a  double layer graphene system to screen
the electrostatic field of an external charge. In the paired state at $T=0$ the
electrostatic field remains completely unscreened at large distances. It was
found\cite{31} that the electron-hole pairing influences significantly the spectrum of
plasma excitations in a double layer graphene system. Namely, instead of one optical
(symmetric) plasmon mode two symmetric modes emerge. The frequency of the lower mode is
restricted from above by the inequality $\hbar\omega<2\Delta$, where $2\Delta$ is the gap
in the electron spectrum caused by the electron-hole pairing. This mode is a weakly
damped one and its frequency is very sensitive to the temperature. At $T=0$ the lower
mode disappears. In contrast, the upper mode belongs to the frequency domain
$\hbar\omega>2\Delta$, it is strongly damped mode, its frequency is less sensitive to the
temperature and it survives at $T=0$. It was also established\cite{32} that the
electron-hole pairing provokes a huge increase of efficiency of the third-harmonic
generation in double monolayer and double bilayer graphenes.

The results \cite{30,31,32}  were obtained within an approach that does not account for
the oscillations of the order parameter of the electron-hole pairing. It is known from
the Bardeen-Cooper-Schrieffer (BCS)  theory of superconductivity \cite{33,34} that
neglecting the order parameter oscillations results in a violation of the gauge
invariancy of the polarization matrix. The gauge invariance is restored by ``dressing''
of the vertexes. The ``dressed'' vertexes should satisfy the generalized Ward identity.
In Ref. \onlinecite{31} we  proposed a heuristic approach to the problem.  We  obtained
the gauge invariant polarization matrix using the vertex functions obtained as particular
solutions of  the generalized Ward identity.

In this paper we present an approach in which the order parameter oscillations are
accounted for explicitly. Our approach is close to one developed in Ref. \onlinecite{35}
for
 conventional superconductors.

 In Sec. \ref{s2} we introduce the model  in which the electron-hole pairing is described
 by the order parameter, which is independent of the momenta of paired quasiparticles.
 The perturbation Hamiltonian that accounts for
 the order parameter oscillations is given in Sec. \ref{s3}. In Sec. \ref{s4} the
 analytical expressions for the response functions and the polarization matrix are
 obtained.  In Sec. \ref{s5} we
 derive the dispersion equation  and calculate the eigenmode spectrum. We identify six modes.
 Two modes correspond to in-phase oscillations
 of  the scalar potentials of two layers. It reproduces the result of Ref. \onlinecite{31}.
 Two other modes correspond to out-of-phase oscillations of the scalar potentials
  coupled to in-phase oscillations of two order parameters (two order parameters
  describe pairing in two spin subsystems).
 One of these modes is interpreted as an analog of the Carlson-Goldman mode in superconductors.
 The remaining two modes correspond to out-of-phase oscillations of two order parameters. They
 can be  considered as analogs of the Anderson-Bogoliubov (phase) and Schmid (amplitude)
 modes in neutral superfluids and superconductors.

\section{The model} \label{s2}

We consider  the electron-hole pairing in a double monolayer graphene system where the
concentration of electrons in one layer is equal to the concentration of holes in the
other layer.  {We specify the case of two graphene layers with zero relative twist. The
graphene layers are separated by a dielectric layer with the dielectric constant
$\varepsilon$ and surrounded by a medium with $\varepsilon=1$. The hopping between
graphene layers is neglected.}

We describe the pairing by the order parameter, which is independent of the momentum.
Such an order parameter can be defined self-consistently in the case of contact
interaction between electrons and holes\cite{16,17}. In the model with contact
interaction the problem of finding the collective mode spectrum can be reduced  to a set
of algebraic equations (in the general case for the momentum dependent order parameter
the algebraic equations are transformed into integral ones).

\begin{figure}
\begin{center}
\includegraphics[width=6cm]{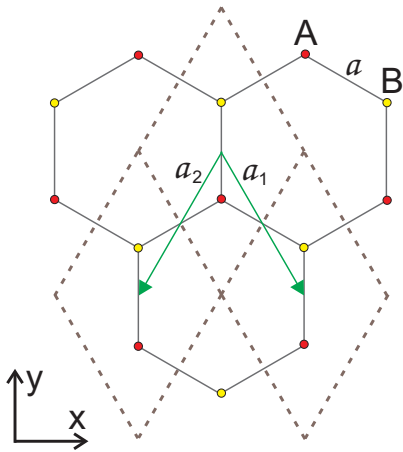}
\end{center}
\caption{ { Graphene lattice. The red (dark) and yellow (light) circles correspond to the
A and B sublattices. The unit cells are shown by dashed lines and two primitive lattice
vectors are shown by arrows.}} \label{f0}
\end{figure}

We describe the system by the Hamiltonian
\begin{equation}\label{1}
    H=H_1+H_2+H_{12},
\end{equation}
where  {
\begin{equation}\label{2}
    H_n=-t\sum_\sigma\sum_i\sum_{j=1,2,3}\left(c^+_{n,i,A,\sigma}c_{n,i+\delta_j,B,\sigma}+
    \mathrm{H.c.}\right)
    -\mu_n\sum_{i,\sigma}\sum_{\alpha}
c^+_{n,i,\alpha,\sigma}c_{n,i,\alpha,\sigma}
\end{equation}
is the single-layer Hamiltonian,  $c^+_{n,i,\alpha,\sigma}$ and $c_{n,i,\alpha,\sigma}$
are the creation and annihilation operators of electrons, $n=1,2$ is the layer index, $i$
is the unit cell index, $\alpha=A,B$ is the sublattice index, $\sigma =
\uparrow,\downarrow$ is the spin index,  $t$ is the nearest-neighbor hopping energy,
$\mu_n$ is the electron chemical potential in the $n$-th layer, and the index
$i+\delta_j$ stands for the unit cell with the coordinate $\mathbf{R}_{i}+\bm{\delta}_j$.
Here  $\mathbf{R}_i$ is the radius-vector of the $i$-th unit cell,  vectors
$(\bm{\delta}_1,\bm{\delta}_2,\bm{\delta}_3)=(0,\mathbf{a}_1,\mathbf{a}_2)$ connect a
given unit cell with unit cells where the nearest-neighbor B sites are located,
 $\mathbf{a}_{1(2)}=(\pm \sqrt{3}a/2,-3a/2)$  are the primitive lattice vectors, and $a$
is the distance between the nearest neighbor atoms in graphene (see Fig. \ref{f0}).}

The chemical potentials are counted from the Dirac points and satisfy the condition
$\mu_1=-\mu_2=\mu$ that corresponds to equal concentrations of electrons and holes.
 The interaction part of the Hamiltonian reads
\begin{equation}\label{3}
    H_{12}=V\sum_{i,\alpha,\sigma}c^+_{1,i,\alpha,\sigma}c^+_{2,i,\alpha,\sigma}c_{2,i,\alpha,\sigma}
    c_{1,i,\alpha,\sigma},
\end{equation}
where $V$ is the interaction constant ($V>0$).

The order parameter of the electron-hole pairing  is defined as
\begin{equation}\label{3-1}
   \Delta_{i,\alpha,\sigma}=V\langle c^+_{2,i,\alpha,\sigma} c_{1,i,\alpha,\sigma}\rangle.
\end{equation}
The order parameter can be presented as a sum of the equilibrium part
$\Delta^{(0)}_{i,\alpha,\sigma}$ and the  fluctuating part
$\Delta^{(fl)}_{i,\alpha,\sigma}(t)$. We consider the paired state with the lowest
energy\cite{16,17} that corresponds to the choice
$\Delta^{(0)}_{i,A,\sigma}=-\Delta^{(0)}_{i,B,\sigma}=\Delta$.  { The property
$\Delta_{i,A}=-\Delta_{i,B}$ provides the opening of the gap in the quasiparticle
spectrum. The contact interaction model with $\Delta_{i,A}=-\Delta_{i,B}$\cite{16,17} and
the model based on a treatment of the long-range Coulomb interaction (bare or screened)
\cite{14,15,18,19} give similar results. In addition, keeping in mind that in the Dirac
approximation the conduction-band and valence-band states are described by the sublattice
spinors $\left(1/\sqrt{2},e^{i\theta_\mathbf{k}}/\sqrt{2}\right)$ and
$\left(1/\sqrt{2},-e^{i\theta_\mathbf{k}}/\sqrt{2}\right)$ respectively
($\hbar\mathbf{k}$ is momentum measured from the Dirac point and $\theta_{\mathbf{k}}$ is
the angular orientation of this momentum) one can see\cite{17} that the order parameter
with $\Delta_{i,A}=-\Delta_{i,B}$ couples the conduction-band and valence-band states
with equal strength at all $\theta_{\mathbf{k}}$.}

Neglecting the order parameter oscillations we obtain the mean-field Hamiltonian
\begin{equation}\label{4}
    H_{MF}=H_1+H_2-\sum_{i,\sigma}\left(\Delta c^+_{1,i,A,\sigma}c_{2,i,A,\sigma}-
    \Delta c^+_{1,i,B,\sigma}c_{2,i,B,\sigma}+\mathrm{H.c.}\right).
\end{equation}
 Applying the Fourier-transformation to
the Hamiltonian (\ref{4})  and considering one spin component we get
\begin{equation}\label{5}
    H_{MF}=\sum_{\mathbf{k}}\Psi^+_\mathbf{k}{h}_\mathbf{k}\Psi_\mathbf{k}=\sum_{\mathbf{k}}
    \left(
             \begin{array}{cccc}
               c^+_{1,A,\mathbf{k}} & c^+_{1,B,\mathbf{k}} & c^+_{2,A,\mathbf{k}} &
               c^+_{2,B,\mathbf{k}} \end{array} \right)   \left(
                                    \begin{array}{cccc}
                                      -\mu & f_\mathbf{k} & -\Delta & 0 \\
                                      f^*_\mathbf{k} & -\mu & 0 & \Delta \\
                                      -\Delta & 0 & \mu & f_\mathbf{k} \\
                                      0 & \Delta & f^*_\mathbf{k} & \mu \\
                                    \end{array}\right)
          \left(                                                \begin{array}{c}
                                                   c_{1,A,\mathbf{k}}  \\
                                                   c_{1,B,\mathbf{k}} \\
                                                   c_{2,A,\mathbf{k}}  \\
                                                   c_{2,B,\mathbf{k}}\\
                                                 \end{array}
                                               \right),\end{equation}
where  $c_{n,A(B),\mathbf{k}}=(1/\sqrt{N})\sum_i c_{n,i,A(B)} e^{-i\mathbf{k}
\mathbf{R}_i}$ is the Fourier-transformed annihilation operator,  $N$ is the total number
of unit cells and the creation operator is given by the Hermitian conjugate, and
$f_\mathbf{k}=|f_\mathbf{k}|e^{i\chi_\mathbf{k}}=-t\sum_{j=1,2,3}\exp(i\mathbf{k}\bm{\delta}_j)$.
Here we omit the spin index.

The Hamiltonian (\ref{5}) is diagonalized by  the unitary transformation
\begin{equation}\label{6}
    H_{MF}=\sum_{\mathbf{k}}\Psi^+_\mathbf{k}\hat{U}^{-1}_\mathbf{k}\hat{U}_\mathbf{k}
{h}_\mathbf{k}\hat{U}^{-1}_\mathbf{k}\hat{U}_\mathbf{k}\Psi_\mathbf{k}
=\sum_{\mathbf{k}}\tilde{\Psi}^+_\mathbf{k}\tilde{h}_\mathbf{k}\tilde{\Psi}_\mathbf{k},
\end{equation}
 where $\tilde{h}_{\mathbf{k}} = \hat{U}_\mathbf{k}{h}_\mathbf{k}\hat{U}^{-1}_\mathbf{k}$
 and $\tilde{\Psi}_\mathbf{k}=\hat{U}_\mathbf{k}\Psi_\mathbf{k}$.
 The matrix $\hat{U}_{\mathbf{k}}$ can be written in a form of the product
\begin{equation}\label{9}
    \hat{U}_\mathbf{k}=\hat{U}_{uv}\hat{U}_{b}\hat{U}_{\chi}.
\end{equation}
The matrix
\begin{equation}\label{10}
    \hat{U}_{\chi}=\frac{1}{\sqrt{2}}\left(
                            \begin{array}{cccc}
                              1 & e^{\mathrm{i}\chi_\mathbf{k}} & 0 & 0 \\
                              1 & -e^{\mathrm{i}\chi_\mathbf{k}}  & 0 & 0 \\
                              0 & 0 & 1 & e^{\mathrm{i}\chi_\mathbf{k}}  \\
                              0 & 0 & 1 & -e^{\mathrm{i}\chi_\mathbf{k}}  \\
                            \end{array}
                          \right)
\end{equation}
diagonalizes the single-layer parts of the Hamiltonian. The matrix
\begin{equation}\label{11}
    \hat{U}_b=\left(
          \begin{array}{cccc}
            1 & 0 & 0 & 0 \\
            0 & 0 & 0 & 1 \\
            0 &1 & 0 & 0 \\
            0 & 0 & 1 & 0 \\
          \end{array}
        \right)
\end{equation}
rearranges the elements of the matrix $\hat{U}_\chi{h}_\mathbf{k}\hat{U}^{-1}_\chi$ into
two blocks:
\begin{equation}\label{12}
    \hat{U}_b\hat{U}_{\chi} {h}_\mathbf{k}\hat{U}^{-1}_{\chi}
    \hat{U}_b^{-1}=\left(
                                         \begin{array}{cccc}
                                           \xi_{\mathbf{k},+1} & -\Delta & 0 & 0 \\
                                           -\Delta & -\xi_{\mathbf{k},+1} & 0 & 0 \\
                                           0 & 0 & \xi_{\mathbf{k},-1} & -\Delta \\
                                           0 & 0 & -\Delta & -\xi_{\mathbf{k},-1} \\
                                         \end{array}
                                     \right),
\end{equation}
where $\xi_{\mathbf{k},\lambda}=\lambda |f_\mathbf{k}|-\mu$ is the electron spectrum of a
single graphene layer,  and  $\lambda=\pm 1$ corresponds to the conduction (valence)
band.

Each block can be diagonalized by the u-v transformation. The matrix
\begin{equation}\label{13}
    \hat{U}_{uv}=\left(
             \begin{array}{cccc}
               u_{\mathbf{k},+1} & -v_{\mathbf{k},+1} & 0 & 0 \\
               v_{\mathbf{k},+1} & u_{\mathbf{k},+1} & 0 & 0 \\
               0 & 0 & u_{\mathbf{k},-1} & -v_{\mathbf{k},-1} \\
               0 & 0 & v_{\mathbf{k},-1} & u_{\mathbf{k},-1} \\
             \end{array}
           \right)
\end{equation}
is expressed through the coefficients of this transformation:
\begin{equation}\label{14}
 u_{\mathbf{k},\lambda}=\sqrt{\frac{1}{2}\left(1+\frac{\xi_{\mathbf{k},\lambda}}
 {E_{\mathbf{k},\lambda}}\right)},
 \quad v_{\mathbf{k},\lambda}=\sqrt{\frac{1}{2}\left(1-\frac{\xi_{\mathbf{k},\lambda}}
 {E_{\mathbf{k},\lambda}}\right)},
\end{equation}
where $E_{\mathbf{k},\lambda}=\sqrt{\xi_{\mathbf{k},\lambda}^2+\Delta^2}$.

The transformed Hamiltonian has the diagonal form:
\begin{equation}\label{15}
H_{MF}=\sum_\nu E_\nu \alpha_\nu^+\alpha_\nu,
\end{equation}
where $\nu=(\mathbf{k},\lambda,m)$ is the full set of the quasiparticle quantum numbers,
excluding spin, $E_\nu=m E_{\mathbf{k},\lambda}$ is the quasiparticle energy,
 {$m=\pm 1$ corresponds to the states above (below) the gap}, and
$\alpha_\nu^+$, $\alpha_\nu$ are the quasiparticle creation and annihilation operators.

Applying to Eq. (\ref{3-1}) the Fourier-transformation and the unitary transformation
$\hat{U}_\mathbf{k}$ we obtain the following equation for the order parameter:
\begin{equation}\label{15-1}
    \Delta=-\frac{V}{2 N}\sum_{\nu} m u_{\mathbf{k},\lambda}v_{\mathbf{k},\lambda}\langle \alpha^+_\nu
    \alpha_{\nu} \rangle.
\end{equation}
Replacing the  average $\langle \alpha^+_\nu
    \alpha_\nu \rangle$ with the Fermi distribution function
     {and calculating the sum over
    $m$},
  we arrive at the self-consistence equation
 \begin{equation}\label{16}
    \Delta=\frac{V \Omega_0}{2 S}\sum_{\mathbf{k},\lambda}\frac{\Delta}
    {2E_{\mathbf{k},\lambda}} \tanh\frac{E_{\mathbf{k},\lambda}}{2T},
\end{equation}
where $\Omega_0$ is the area of the unit cell and $S$ is the area of the system.

We emphasize that Eq. (\ref{16}) differs from one obtained in the model with a long-range
Coulomb interaction \cite{14,15,18,19}. In the latter case the self-consistence equation
has the form
\begin{equation}\label{17}
   \Delta_{\mathbf{k},\lambda}=\frac{1}{S}\sum_{\mathbf{k'},\lambda'}V(\mathbf{k}-\mathbf{k}')
   \frac{1+\lambda\lambda'\cos(\chi_\mathbf{k}-\chi_{\mathbf{k}'})}{2}
   \frac{\Delta_{\mathbf{k}',\lambda'}}{2 E_{\mathbf{k}',\lambda'}}
   \tanh\frac{E_{\mathbf{k'},\lambda'}}{2T},
\end{equation}
where $V(\mathbf{q})$ is the Fourier-component of the interlayer Coulomb interaction.
Differently from Eq. (\ref{16}), the order parameter independent of $\mathbf{k}$ and
$\lambda$  does not satisfy Eq. (\ref{17}).

\section{Perturbation Hamiltonian} \label{s3}

Now we add to the Hamiltonian (\ref{5}) the perturbation part $H_{int}$.  The
perturbation Hamiltonian $H_{int}$ describes the oscillations of the order parameter and
the interaction of electrons with the scalar potential $\varphi(\mathbf{r},t)$. We
consider the oscillations for which
$\Delta^{(fl)}_{i,A,\sigma}(t)=-\Delta^{(fl)}_{i,B,\sigma}(t)=\Delta^{(fl)}_{i,\sigma}(t)$
and do not take into account oscillations with
$\Delta^{(fl)}_{i,A,\sigma}=+\Delta^{(fl)}_{i,B,\sigma}$. The latter ones are decoupled
from the scalar potential oscillations and do not modify the response to the
electromagnetic field.

The Fourier-components of the real and imaginary parts of the order parameter
oscillations are defined as
\begin{equation}\label{20}
    \Delta_1(\mathbf{q},\omega)=\Omega_0\sum_i\int dt e^{\mathrm{i}\omega
    t-\mathrm{i}\mathbf{q}\mathbf{R}_i}\mathrm{Re}[\Delta^{(fl)}_i(t)],
\end{equation}
\begin{equation}\label{21}
    \Delta_2(\mathbf{q},\omega)=\Omega_0\sum_i\int dt e^{\mathrm{i}\omega
    t-\mathrm{i}\mathbf{q}\mathbf{R}_i}\mathrm{Im}[\Delta^{(fl)}_i(t)].
\end{equation}

We specify the case of  real-valued $\Delta$ (it is accounted for in the Hamiltonian
(\ref{5}) and in the coefficients (\ref{14})). Then the quantities $\Delta_1$ and
$\Delta_2$ describe small oscillations of the amplitude and the phase of the order
parameter, respectively.

The perturbation Hamiltonian can be presented in the matrix form
\begin{equation}\label{23}
    H_{int}(t)=-\frac{1}{2\pi S}\sum_{\mathbf{k},\mathbf{q}}\int d \omega e^{-\mathrm{i}\omega t}
    \Psi^+_{\mathbf{k}+\mathbf{q}}\left[\frac{e}{2}\varphi_+(\mathbf{q},\omega)\hat{T}^{(0)}+
    \Delta_1(\mathbf{q},\omega)\hat{T}^{(1)}+
 \Delta_2(\mathbf{q},\omega)\hat{T}^{(2)}+\frac{e}{2}\varphi_-(\mathbf{q},\omega)\hat{T}^{(3)}
    \right]\Psi_{\mathbf{k}},
\end{equation}
where the operators $\Psi^+_\mathbf{k}$ and $\Psi_\mathbf{k}$ are defined by Eq.
(\ref{5}),
\begin{equation}\label{22}
    \varphi_\pm(\mathbf{q},\omega)=\Omega_0\sum_i\int dt e^{\mathrm{i}\omega
    t-\mathrm{i}\mathbf{q}\mathbf{R}_i}\left[\varphi_1(\mathbf{R}_i,t)\pm
    \varphi_2(\mathbf{R}_i,t)\right]
\end{equation}
is the Fourier-component of the sum (difference) of the scalar potentials in two graphene
layers, and $\varphi_{n}(\mathbf{R}_i,t)$ is the scalar potential in the $n$-th layer in
the $i$-th unit cell. The matrices $\hat{T}^{(s)}$ in Eq. (\ref{23}) are expressed
through the Pauli matrix $\hat{\sigma}_z$ and the identity matrix $\hat{I}$:
\begin{equation}\label{24}
\hat{T}^{(0)}=\left(
            \begin{array}{cc}
              \hat{I} & 0 \\
              0 & \hat{I} \\
            \end{array}
          \right), \quad
\hat{ T}^{(1)}=\left(
                  \begin{array}{cc}
                    0 & \hat{\sigma}_z \\
                    \hat{\sigma}_z & 0 \\
                  \end{array} \right),
          \quad
 \hat{T}^{(2)}=\left(
            \begin{array}{cc}
              0 & i\hat{\sigma}_z \\
              -i \hat{\sigma}_z & 0 \\
            \end{array}
          \right), \quad
          \hat{T}^{(3)}=\left(
            \begin{array}{cc}
              \hat{I} & 0 \\
              0 & -\hat{I} \\
            \end{array} \right).
          \end{equation}

We apply the transformation (\ref{6}) to the Hamiltonian (\ref{23}) and write it through
the operators of creation and annihilation of quasiparticle excitations:
\begin{eqnarray}\label{25}
    H_{int}(t)=\frac{1}{2\pi S}\sum_{\nu_1,\nu_2}\int d \omega e^{-\mathrm{i}\omega t}
    \alpha^+_{\nu_1}[h_{int}(\omega)]_{\nu_1,\nu_2}\alpha_{\nu_2},
\end{eqnarray}
where
\begin{equation}\label{25-1}
    [h_{int}(\omega)]_{\nu_1,\nu_2}=-\frac{e}{2}\varphi_+(\mathbf{k}_2-\mathbf{k}_1,\omega)
    R^{(0)}_{\nu_1,\nu_2}-
    \Delta_1(\mathbf{k}_2-\mathbf{k}_1,\omega)R^{(1)}_{\nu_1,\nu_2}-
 \Delta_2(\mathbf{k}_2-\mathbf{k}_1,\omega)R^{(2)}_{\nu_1,\nu_2}
  -\frac{e}{2}\varphi_-(\mathbf{k}_2-\mathbf{k}_1,\omega)R^{(3)}_{\nu_1,\nu_2},
\end{equation}
the matrices $R^{(s)}_{\nu_1,\nu_2}$  ($s=0,1,2,3$) are given by the equation
\begin{equation}\label{26}
 R^{(s)}_{\mathbf{k}_1,\lambda_1,m_1;\mathbf{k}_2,\lambda_2,m_2}
 =\frac{1+\lambda_1\lambda_2e^{\mathrm{i}(\chi_{\mathbf{k}_1}-\chi_{\mathbf{k}_2})}}{2}[M^{(s)}
 (\mathbf{k}_1,\lambda_1,\mathbf{k}_2,\lambda_2)]_{i_{m_1},i_{m_2}},
 \end{equation}
($i_{+1} \equiv 1$,
  $i_{-1} \equiv 2$), and the matrices $\hat{M}^{(s)}$ are expressed through the product
\begin{eqnarray}
\label{27} \hat{M}^{(s)}(\mathbf{k}_1,\lambda_1,\mathbf{k}_2,\lambda_2)=
\left(
  \begin{array}{cc}
    u_{\mathbf{k}_1,\lambda_1} & - v_{\mathbf{k}_1,\lambda_1}\\
v_{\mathbf{k}_1,\lambda_1} & u_{\mathbf{k}_1,\lambda_1} \\
  \end{array}\right)
\hat{\sigma}^{(s)} \left(
  \begin{array}{cc}
    u_{\mathbf{k}_2,\lambda_2} &  v_{\mathbf{k}_2,\lambda_2}\\
-v_{\mathbf{k}_2,\lambda_2} & u_{\mathbf{k}_2,\lambda_2} \\
  \end{array}\right)
\end{eqnarray}
with $\hat{\sigma}^{(0)}=\hat{I}$, $\hat{\sigma}^{(1)}=\hat{\sigma}_x$,
 $\hat{\sigma}^{(2)}=-\hat{\sigma}_y$, $\hat{\sigma}^{(3)}=\hat{\sigma}_z$.

\section{Polarization matrix} \label{s4}

Taking into account two spin components we write the Hamiltonian in the form
\begin{equation}\label{33-1}
    H(t)=H_{MF}+H_{int}(t)=\sum_{\nu,\sigma}E_\nu \alpha^+_{\nu,\sigma}\alpha_{\nu,\sigma}+
\frac{1}{2\pi S}\sum_{\nu_1,\nu_2,\sigma}\int d \omega e^{-\mathrm{i}\omega t}
    \alpha^+_{\nu_1,\sigma}[h_{int,\sigma}(\omega)]_{\nu_1,\nu_2}\alpha_{\nu_2,\sigma},
\end{equation}
where  $[h_{int,\sigma}(\omega)]_{\nu_1,\nu_2}$ is given by Eq. (\ref{25-1}) with
$\Delta_{1(2)}(\mathbf{k},\omega)\equiv \Delta_{1(2),\sigma}(\mathbf{k},\omega)$.

To calculate the response of the system to the scalar potential and to the order
parameter oscillations we define the response functions
\begin{equation}\label{33-2}
   \eta^{(s)}_{\sigma}(\mathbf{q},\omega)=\int dt e^{i\omega t}\sum_{\mathbf{k}}
\langle \Psi_{\mathbf{k}-\mathbf{q},\sigma}^+\hat{
T}^{(s)}\Psi_{\mathbf{k},\sigma}\rangle,
\end{equation}
where $\Psi^+_{\mathbf{k},\sigma}$ and $\Psi_{\mathbf{k},\sigma}$ are the same operators
as in Eq. (\ref{5}) with restored spin indexes. The angle brackets mean the quantum
mechanical and thermodynamic average. We compute the averages in  Eq. (\ref{33-2}) using
the density matrix formalism. The density matrix $\hat{\rho}(t)$ satisfies the equation
\begin{equation}\label{33-21}
    \frac{\partial \hat{\rho}(t)}{\partial
t}=\frac{1}{i\hbar}[H(t),\hat{\rho}(t)]-\gamma(\hat{\rho}(t)-\hat{\rho}_0),
\end{equation}
where $\hat{\rho}_0$ is the density matrix of the system described by the  Hamiltonian
$H_{MF}$, and $\gamma$ is the relaxation rate.  {The quantity $\gamma$ is the
phenomenological parameter. In what follows we consider  small $\gamma$ ($\hbar\gamma\ll
\mu$). It corresponds to the pure limit. Accounting for the term with $\gamma$ in Eq.
(\ref{33-21}) allows to calculate numerically the integrals in the expressions for the
polarization matrix and to evaluate the Landau damping.}

The averages in Eq. (\ref{33-2}) are calculated as
\begin{equation}\label{33-3}
\langle \Psi_{\mathbf{k}-\mathbf{q},\sigma}^+\hat{
T}^{(s)}\Psi_{\mathbf{k},\sigma}\rangle
=\mathrm{Tr}\left([\hat{\rho}(t)]_{\mathbf{k},\sigma; \mathbf{k}-\mathbf{q},\sigma} \hat{
T}^{(s)}\right),
\end{equation}
 where the trace  is taken over the sublattice and layer indexes.

In the quasiparticle  basis the response functions (\ref{33-2}) are expressed as
\begin{equation}\label{36}
\eta^{(s)}_{\sigma}(\mathbf{q},\omega)=\sum_{\nu_1,\nu_2}
[\hat{\rho}(\omega)]_{\nu_1,\sigma; \nu_2,\sigma}
R^{(s)}_{\nu_2,\nu_1}\delta_{\mathbf{k}_1-\mathbf{q},\mathbf{k}_2},
\end{equation}
where $\hat{\rho}(\omega)=\int d t \exp(\mathrm{i}\omega t)\hat{\rho}(t)$ and the
matrixes $R^{(s)}_{\nu_1,\nu_2}$ are given by Eq. (\ref{26}).

  The density matrix is sought in a form of expansion in powers of the perturbation Hamiltonian:
$\hat{\rho}(\omega)=\hat{\rho}_0(\omega)+\hat{\rho}_1(\omega)+\ldots$. The zero order
term in this expansion is the equilibrium density matrix
\begin{equation}\label{36-1}
[\hat{\rho}_0(\omega)]_{\nu_1,\sigma_1; \nu_2,\sigma_2}=2\pi\delta(\omega)\delta_{\nu_1,\nu_2}
\delta_{\sigma_1,\sigma_2} f_{\nu_1},
\end{equation}
where $f_\nu=(e^{E_\nu/T}+1)^{-1}$ is the Fermi distribution function. The first order
term reads
\begin{equation}\label{37}
    [\hat{\rho}_1(\omega)]_{\nu_1,\sigma_1; \nu_2,\sigma_2}=\frac{1}{S}\frac{f_{\nu_1}-f_{\nu_2}}
    {E_{\nu_1}-E_{\nu_2}-\hbar(\omega+\mathrm{i}\gamma)} [h_{int,\sigma_1}(\omega)]_{\nu_1,\nu_2}
\delta_{\sigma_1,\sigma_2}.
\end{equation}

The response functions $\eta^{(0)}$ and $\eta^{(3)}$ at $\mathbf{q}\ne 0$ correspond to
the charge density oscillations $\rho_{\pm,\sigma}=\rho_{1,\sigma}\pm \rho_{2,\sigma}$:
\begin{equation}\label{37-1}
\rho_{+,\sigma}(\mathbf{q},\omega)=-e \eta^{(0)}_{\sigma}(\mathbf{q},\omega), \quad
\rho_{-,\sigma}(\mathbf{q},\omega)=-e \eta^{(3)}_{\sigma}(\mathbf{q},\omega).
\end{equation}
Taking into account the definition of the order parameter Eq. (\ref{3-1}) we obtain the
relation between the order parameter oscillations and the response functions
$\eta^{(1(2))}$ at $\mathbf{q}\ne 0$:
\begin{equation}\label{37-2}
\Delta_{1,\sigma}(\mathbf{q},\omega)=g\eta^{(1)}_{\sigma}(\mathbf{q},\omega), \quad
\Delta_{2,\sigma}(\mathbf{q},\omega)=g\eta^{(2)}_{\sigma}(\mathbf{q},\omega),
\end{equation}
where $g=V\Omega_0/4$ is the coupling constant.

Restricting with the linear response approximation we obtain from Eqs. (\ref{36}),
(\ref{37}), (\ref{37-1}), and (\ref{37-2}) the following matrix equation
\begin{equation}\label{38}
    \left(
      \begin{array}{c}
      e^{-1}\rho_{+,\sigma}(\mathbf{q},\omega) \\
       -g^{-1}\Delta_{1,\sigma}(\mathbf{q},\omega) \\
       -g^{-1} \Delta_{2,\sigma}(\mathbf{q},\omega) \\
        e^{-1}\rho_{-,\sigma}(\mathbf{q},\omega) \\
      \end{array}
    \right)=\left(
              \begin{array}{cccc}
                \Pi_{00}(\mathbf{q},\omega) & \Pi_{01}(\mathbf{q},\omega) &
                \Pi_{02}(\mathbf{q},\omega) & \Pi_{03}(\mathbf{q},\omega) \\
                \Pi_{10}(\mathbf{q},\omega) & \Pi_{11}(\mathbf{q},\omega) &
                \Pi_{12}(\mathbf{q},\omega) & \Pi_{13}(\mathbf{q},\omega)  \\
                \Pi_{20}(\mathbf{q},\omega) & \Pi_{21}(\mathbf{q},\omega) &
                \Pi_{22}(\mathbf{q},\omega) & \Pi_{23}(\mathbf{q},\omega)\\
                \Pi_{30}(\mathbf{q},\omega) & \Pi_{31}(\mathbf{q},\omega) &
                \Pi_{32}(\mathbf{q},\omega) & \Pi_{33}(\mathbf{q},\omega)\\
              \end{array}
            \right)\left(
                     \begin{array}{c}
                       e\varphi_+(\mathbf{q},\omega)/2\\
                       \Delta_{1,\sigma}(\mathbf{q},\omega)\\
                       \Delta_{2,\sigma}(\mathbf{q},\omega) \\
                       e\varphi_-(\mathbf{q},\omega)/2\\
                     \end{array}
                   \right) ,
\end{equation}
where  the components of the polarization matrix are given by the expression
\begin{equation}\label{39}
\Pi_{s_1s_2}(\mathbf{q},\omega)=\frac{1}{S}\sum_{\nu_1,\nu_2}
\delta_{\mathbf{k}_1-\mathbf{q},\mathbf{k}_2} \Phi^{s_1 s_2}_{\nu_1\nu_2}
\frac{1+\lambda_1\lambda_2\cos(\chi_{\mathbf{k}_1}-\chi_{\mathbf{k}_2})}{2}\frac{f_{\nu_1}-f_{\nu_2}}
    {E_{\nu_1}-E_{\nu_2}-\hbar(\omega+i\gamma)}. \end{equation}
 The factors $\Phi^{s_1 s_2}_{\nu_1\nu_2}$ in Eq. (\ref{39}) are expressed through the matrix
(\ref{27}):
\begin{equation}\label{40}
   \Phi^{s_1
   s_2}_{\nu_1\nu_2}=[\hat{M}^{(s_2)}(\mathbf{k}_1,\lambda_1,\mathbf{k}_2,\lambda_2)]_{i_{m_1},i_{m_2}}
   [\hat{M}^{(s_1)}(\mathbf{k}_2,\lambda_2,\mathbf{k}_1,\lambda_1)]_{i_{m_2},i_{m_1}},
\end{equation}
(there is no summation over repeated indexes in Eq. (\ref{40})).

From Eq. (\ref{40}) we obtain the following explicit expressions for  $\Phi^{s_1
s_2}_{\nu_1\nu_2}$ :
  \begin{eqnarray} \label{41}
    \Phi^{00}_{\nu_1\nu_2} = \frac{1}{2}\left(1+\frac{\xi_1\xi_2+\Delta^2}{E_1
    E_2}\right), \quad
 \Phi^{01}_{\nu_1\nu_2} =-\frac{\Delta}{2}
    \left(\frac{1}{E_1}+\frac{1}{E_2}\right), \quad
    \Phi^{02}_{\nu_1\nu_2}
    =i\frac{\Delta}{2}\frac{\xi_2-\xi_1}{E_2E_1}, \quad
    \Phi^{03}_{\nu_1\nu_2}      =\frac{1}{2}\left(\frac{\xi_2}{E_2}+\frac{\xi_1}{E_1}\right), \cr
     \Phi^{11}_{\nu_1\nu_2}=\frac{1}{2}\left(1-\frac{\xi_1\xi_2-\Delta^2}{E_1
     E_2}\right),\quad
\Phi^{12}_{\nu_1\nu_2} = \frac{i}{2}\left(\frac{\xi_1}{E_1}
    -\frac{\xi_2}{E_2}\right), \quad
    \Phi^{13}_{\nu_1\nu_2} =
    -\frac{\Delta}{2}\frac{\xi_1+\xi_2}{E_1E_2}, \cr
    \Phi^{22}_{\nu_1\nu_2} = \frac{1}{2}\left(1-\frac{\xi_1\xi_2+\Delta^2}{E_1 E_2}\right),
    \quad
\Phi^{23}_{\nu_1\nu_2}
    =i\frac{\Delta}{2}\left(\frac{1}{E_2}-\frac{1}{E_1}\right),\cr
    \Phi^{33}_{\nu_1\nu_2} = \frac{1}{2}\left(1+\frac{\xi_1\xi_2-\Delta^2}{E_1 E_2}\right)
\end{eqnarray}
and $\Phi^{s_2,s_1}_{\nu_1\nu_2}=(\Phi^{s_1,s_2}_{\nu_1\nu_2})^*$. Here we use the
notations $\xi_i\equiv\xi_{\nu_i}$ and $E_i\equiv E_{\nu_i}$.

Taking into account symmetry properties of the expression under summation in Eq.
(\ref{39}), one can show that some elements of the polarization matrix, namely,
$\Pi_{01}(\mathbf{q},\omega)$, $\Pi_{02}(\mathbf{q},\omega)$,
$\Pi_{03}(\mathbf{q},\omega)$, $\Pi_{10}(\mathbf{q},\omega)$,
$\Pi_{20}(\mathbf{q},\omega)$, and $\Pi_{30}(\mathbf{q},\omega)$, are equal to zero
exactly.

\section{Collective modes} \label{s5}

 In the nonretarded
 approximation the scalar potential satisfies the Poisson equation
\begin{equation}\label{44}
    \nabla[\varepsilon(\mathbf{r})\nabla\varphi(\mathbf{r},t)]=-4\pi\rho(\mathbf{r},t),
\end{equation}
where
\begin{equation}\label{45}
    \varepsilon(\mathbf{r})=\left\{
                     \begin{array}{ll}
                       1, & \hbox{$z<-d/2$;} \\
                       \varepsilon, & \hbox{$-d/2<z<d/2$;} \\
                       1, & \hbox{$z>d/2$,}
                     \end{array}
                   \right.
\end{equation}
is the space-dependent dielectric constant (we specify the case of two graphene layers
separated by a dielectric layer with the dielectric constant $\varepsilon$ and surrounded
by a medium with $\varepsilon=1$), $d$ is the distance between graphene layers, and the
$z$-axis is directed perpendicular to graphene layers.

To obtain the eigenmode spectrum we account for  the  charges
 induced in graphene layers  by the scalar potential and by the
order parameter oscillations in Eq. (\ref{44}):
\begin{equation}\label{46}
\rho(\mathbf{r},t)=\sum_\sigma[\rho_{1,\sigma}(\mathbf{r}_{pl},t)\delta(z-d/2)+\rho_{2,\sigma}
(\mathbf{r}_{pl},t)\delta(z+d/2)],
\end{equation}
where $\mathbf{r}_{pl}$ is two-dimensional radius-vector in the $(x,y)$-plane.

Making the Fourier-transformation of  Eq. (\ref{44}) we obtain the equation for
$\varphi(\mathbf{q},z,\omega)$. Its solution yields the relation between the potentials
$\varphi_\pm(\mathbf{q},\omega)=\varphi(\mathbf{q},d/2,\omega)\pm
\varphi(\mathbf{q},-d/2,\omega)$ and the charge densities
$\rho_{\pm}(\mathbf{q},\omega)=\sum_\sigma \rho_{\pm,\sigma}(\mathbf{q},\omega)$:
\begin{equation}\label{47}
    e^2\varphi_{\pm}(\mathbf{q},\omega)=V_{\pm}(q)\rho_{\pm}(\mathbf{q},\omega),
\end{equation}
where
\begin{equation}\label{48}
    V_{\pm}(q)=\frac{4\pi e^2}{q}\frac{1\pm e^{-q d}}{(\varepsilon+1)\mp (\varepsilon-1)e^{- q d}}
\end{equation}
are the Fourier-components of the Coulomb interaction energies
$V_{\pm}(r_{pl})=V_{11}(r_{pl})\pm V_{12}(r_{pl})$. Here  $V_{11}(r_{pl})$ and
$V_{12}(r_{pl})$ are the energies of interaction of two electrons located in the same and
different layers, correspondingly (we account for that in the uniform dielectric
environment $V_{11}(r_{pl})=V_{22}(r_{pl})$).

From Eqs. (\ref{38}) and (\ref{47}) we get the equation for the scalar potential and
order parameter oscillations:
\begin{equation}\label{49}
    \left(
      \begin{array}{cccccc}
        2\Pi_{00}- \frac{2}{V_+(q)}& 0 & 0 & 0 & 0 & 0 \\
        0 & \Pi_{11}+\frac{1}{g} & \Pi_{12} & 0 & 0 & \Pi_{13} \\
        0 & \Pi_{21} & \Pi_{22}+\frac{1}{g} & 0 & 0 & \Pi_{23} \\
        0 & 0 & 0 &  \Pi_{11}+\frac{1}{g}& \Pi_{12} & \Pi_{13} \\
        0 & 0 & 0 & \Pi_{21} & \Pi_{22}+\frac{1}{g} & \Pi_{23} \\
        0 & \Pi_{31} & \Pi_{32} & \Pi_{31} & \Pi_{32} & 2\Pi_{33}-\frac{2}{V_-(q)} \\
      \end{array}
    \right)\left(
             \begin{array}{c}
                e\varphi_+(\mathbf{q},\omega)/2 \\
               \Delta_{1,\uparrow}(\mathbf{q},\omega) \\
               \Delta_{2,\uparrow}(\mathbf{q},\omega) \\
               \Delta_{1,\downarrow}(\mathbf{q},\omega)\\
                \Delta_{2,\downarrow}(\mathbf{q},\omega) \\
                e\varphi_-(\mathbf{q},\omega)/2 \\
             \end{array}
           \right)=0,
\end{equation}
where $\Pi_{\alpha\beta}\equiv\Pi_{\alpha\beta}(\mathbf{q},\omega)$.

We calculate the polarization functions Eq. (\ref{39}) in the Dirac approximation for the
electron spectrum.  In this approximation the sum over $\mathbf{k}$ is replaced with the
integral over two separate circles in the Brillouin zone centered at the Dirac points $K$
and $K'$. In these circles $|f(\mathbf{k})| \approx \hbar v_F k'$, and $\chi_{\mathbf{k}}
\approx \mp \theta_{\mathbf{k}'}$, where $\mathbf{k}'$ is counted from the corresponding
Dirac point, $\theta_{\mathbf{k}'}$ is the angle between $\mathbf{k}'$ and the $x$-axis,
and $v_F$ is the Fermi velocity in graphene.  { In the Dirac approximation the integrals
in the expressions for $\Pi_{11}(\mathbf{q},\omega)$ and $\Pi_{22}(\mathbf{q},\omega)$
diverge at $k'\to \infty$. This divergence is unphysical one and emerges as a result of
the approximations used. The same (unphysical) divergence emerges in the self-consistence
equation (\ref{16}) if it is evaluated in the Dirac approximation. Fortunately the
quantities $\Pi_{11(22)}(\mathbf{q},\omega)+1/g$ that enter into Eq. (\ref{49}) can be
presented in a form that is free from such a divergence. Indeed Eq. (\ref{16}) can be
rewritten as}
\begin{equation}\label{42-1}
 \frac{1}{g}=\frac{1}{S}\sum_{\mathbf{k},\lambda}
 \frac{1}{E_{\mathbf{k},\lambda}} \tanh\frac{E_{\mathbf{k},\lambda}}{2T}=-
\frac{1}{S}\sum_{m,\mathbf{k},\lambda}
\frac{f_{m,\mathbf{k},\lambda}-f_{-m,\mathbf{k},\lambda}}
{E_{m,\mathbf{k},\lambda}-E_{-m,\mathbf{k},\lambda}}
\end{equation}
Using the relation (\ref{42-1}) we get
 \begin{eqnarray}\label{43}
\Pi_{ss}(\mathbf{q},\omega)+\frac{1}{g}= \Pi_{ss}^{(R)}(\mathbf{q},\omega)\cr=
\frac{1}{S}\sum_{\nu_1,\nu_2} \left[\delta_{\mathbf{k}_1-\mathbf{q},\mathbf{k}_2}
\Phi^{ss} _{\nu_1\nu_2}
\frac{1+\lambda_1\lambda_2\cos(\chi_{\mathbf{k}_1}-\chi_{\mathbf{k}_2})}{2}\frac{f_{\nu_1}-f_{\nu_2}}
    {E_{\nu_1}-E_{\nu_2}-\hbar(\omega+i\gamma)}-\delta_{\mathbf{k}_1,\mathbf{k}_2}
\delta_{m_1,-m_2}\delta_{\lambda_1,\lambda_2}\frac{f_{\nu_1}-f_{\nu_2}}
    {E_{\nu_1}-E_{\nu_2}}\right].
\end{eqnarray}
 {that do not diverge in the Dirac approximation (divergencies in
$\Pi_{11(22)}$ and in Eq.( \ref{42-1}) cancel each other).}

Equating the determinant of the matrix in Eq. (\ref{49}) to zero we obtain the dispersion
equation for the eigenmode spectrum. The determinant is factorized into three
multipliers. The first multiplier yields the equation
\begin{equation}\label{50}
    \varepsilon_+(\mathbf{q},\omega)=1-V_+(q)\Pi_{00}(\mathbf{q},\omega)=0.
\end{equation}
Equation (\ref{50})  is the dispersion equation for the symmetric plasma excitation in
the double layer system.

The dielectric function $\varepsilon_+(\mathbf{q},\omega)$ describes the screening of the
scalar potential of a test charge $\rho^{\mathrm{test}}_+(\mathbf{q},\omega)$:
  \begin{equation}\label{53}
    e^2\varphi^{\mathrm{scr}}_+(\mathbf{q},\omega)=\frac{V_{+}(q)}{\varepsilon_+(\mathbf{q},\omega)}
\rho^{\mathrm{test}}_+(\mathbf{q},\omega).
  \end{equation}
Equation (\ref{53}) follows from Eq. (\ref{47}) written in the form $e^2
\varphi^{\mathrm{scr}}_+(\mathbf{q},\omega)=V_{+}(q)[\rho^{\mathrm{test}}_+(\mathbf{q},\omega)+
\rho^{\mathrm{ind}}_+(\mathbf{q},\omega)]$, where
$\rho^{\mathrm{ind}}_+(\mathbf{q},\omega)=e^2
\Pi_{00}(\mathbf{q},\omega)\varphi^{\mathrm{scr}}_+(\mathbf{q},\omega)$ is the induced
charge.

From the continuity equation for the charge we obtain the relation between the
polarization function $\Pi_{00}(\mathbf{q},\omega)$ and the longitudinal parallel current
conductivity $\sigma_{+,xx}(\mathbf{q},\omega)$:
\begin{equation}\label{55}
    \sigma_{+,xx}(q \mathbf{i}_x,\omega)
=\frac{ie^2\omega}{q^2}\Pi_{00}(q \mathbf{i}_x,\omega),
\end{equation}
where $\mathbf{i}_{x}$ is the unit vector along the x axis.

Considering the Maxwell's equations with the corresponding boundary conditions and the
matter equation for the current one can get the following dispersion equation for the
symmetric plasmon modes \cite{31,36}:
\begin{equation}\label{56}
    1+\frac{4\pi i \kappa_1}{\omega}\sigma_{+,xx}(q \mathbf{i}_x,\omega)
+\frac{\varepsilon \kappa_1}{\kappa_2}\tanh\frac{\kappa_2 d}{2}=0,
\end{equation}
where $\kappa_1=\sqrt{q^2-\omega^2/c^2}$ and $\kappa_2=\sqrt{q^2-\varepsilon
\omega^2/c^2}$.  Equation (\ref{56}) accounts for retarded effects, and due to this it
differs from Eq. (\ref{50}). In the limit $\kappa_1=\kappa_2=q$ that corresponds to
nonretarded (plasmon) approximation Eq. (\ref{56}) is reduced to Eq. (\ref{50}).

Thus, we have shown that the order parameter oscillations are decoupled from the
oscillations of $\varphi_+$ and do not influence the spectrum of symmetric plasmon modes.
The same result was obtained in Ref. \onlinecite{31} based on the observation that the
generalized Ward identity for the vertex function $\Gamma_{\mu,+}$ is satisfied with bare
vertexes (the vertexes $\Gamma_{\mu,+}$ describe interaction of electrons with
$\varphi_+$ and $\mathbf{A}_+$, the sum of vector potentials of two layers). Therefore
the Feynman diagram with the bare vertexes (which do not account for order parameter
oscillations) gives a gauge invariant polarization function $\Pi_{00}$.
 {The gauge invariance of $\Pi_{00}$ can be also checked directly (see the
Appendix)}.

 In the general case Eq. (\ref{50}) has two
solutions\cite{31}, one is below the gap ($\hbar\omega<2\Delta$), and the other is above
the gap ($\hbar\omega>2\Delta$). It can be seen from the frequency dependence of the
dielectric loss function. This function is defined as
 \begin{equation}\label{52}
    L_+(\mathbf{q},\omega)=-\mathrm{Im}\left[\frac{1}{\varepsilon_+(\mathbf{q},\omega)}\right].
 \end{equation}
It determines relative losses of energy of oscillations of a test charge $\rho_+^{test}$.
The positions of peaks in the $\omega$-dependence of $L_+(q,\omega)$ at fixed $q$
correspond to the eigenmode frequencies. A half-width of the peak at its half-height
gives the damping rate for the corresponding mode.

\begin{figure}
\begin{center}
\includegraphics[width=8cm]{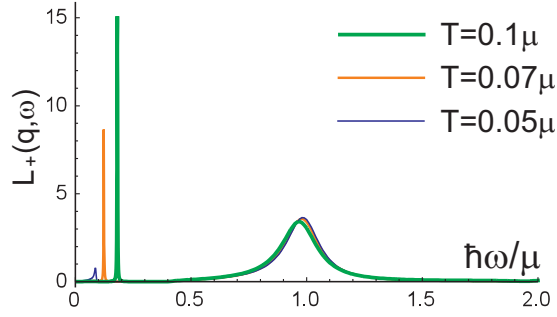}
\end{center}
\caption{Frequency dependence of the dielectric loss function (\ref{52}) at $T=0.1\mu,\
0.07\mu,\ 0.05\mu$, $q=0.1 k_F$, $\Delta=0.2\mu$, and $\gamma=10^{-3}\mu$} \label{f1}
\end{figure}

\begin{figure}
\begin{center}
\includegraphics[width=8cm]{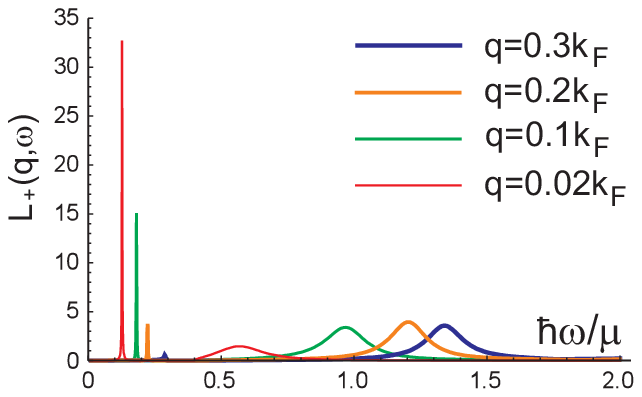}
\end{center}
\caption{Frequency dependence of the dielectric loss function (\ref{52}) at $T=0.1\mu$,
$q=0.02 k_F,\ 0.1 k_F,\ 0.2 k_F,\ 0.3 k_F$, $\Delta=0.2\mu$, and $\gamma=10^{-3}\mu$}
\label{f2}
\end{figure}

To compare the properties of symmetric and antisymmetric (see below) modes it is
instructive to illustrate changes in the frequency dependence of $L_+(\mathbf{q},\omega)$
under variation of temperature (Fig. \ref{f1}) and the wave vector (Fig. \ref{f2}). One
can see that the peak that corresponds to the lower mode disappears at small $T$ and for
large $q$.   One can also see in Figs. \ref{f1} and \ref{f2} a wide peak that corresponds
to the upper (strongly damped) mode. Note that at $\Delta\to 0$ the damping rate of the
upper mode decreases and this mode is transformed into the normal state optical plasmon
mode.

The second  multiplier in the determinant of the matrix in Eq. (\ref{49}) yields the
equation
\begin{equation}\label{57}
\Pi_{11}^{(R)}(\mathbf{q},\omega)\Pi_{22}^{(R)}(\mathbf{q},\omega)
+[\Pi_{12}(\mathbf{q},\omega)]^2=0.
\end{equation}
 It is the dispersion equation for the excitations where only the difference
$\Delta_\uparrow -\Delta_\downarrow$ oscillates. Such oscillations are decoupled from the
scalar potential oscillations.

In the theory of superconductivity  the eigenmodes that correspond to oscillations of the
phase and the modulus of the order parameter are known as the Anderson-Bogoliubov (AB)
mode\cite{37,38} and the Schmid \cite{39} mode. Since in common superconductors the
oscillations of the phase of the order parameter are coupled to plasma (scalar potential)
oscillations, a genuine Anderson-Bogoliubov mode can emerge in neutral Fermi superfluids.
In double layer systems with electron-hole pairing the presence of two superconducting
components  allows us to realize the AB mode. To visualize the AB and the Schmid modes we
introduce the functions
\begin{equation}\label{58-1}
    L_{11}(\mathbf{q},\omega)=\frac{1}{g}\mathrm{Im}\left[\frac{1}
{\Pi_{11}^{(R)}(\mathbf{q},\omega)+\frac{[\Pi_{12}(\mathbf{q},\omega)]^2}
{\Pi_{22}^{(R)}(\mathbf{q},\omega)}}\right],
\end{equation}
\begin{equation}\label{58-2}
    L_{22}(\mathbf{q},\omega)=\frac{1}{g}\mathrm{Im}\left[\frac{1}
{\Pi_{22}^{(R)}(\mathbf{q},\omega)+\frac{[\Pi_{12}(\mathbf{q},\omega)]^2}
{\Pi_{11}^{(R)}(\mathbf{q},\omega)}}\right].
\end{equation}
These functions can be interpreted as analogs of the energy loss function (\ref{52}). The
functions $L_{11}$ and $L_{22}$ describe losses of energy  under externally driven
oscillations of the amplitude and the phase of the order parameter, respectively.

\begin{figure}
\begin{center}
\includegraphics[width=8cm]{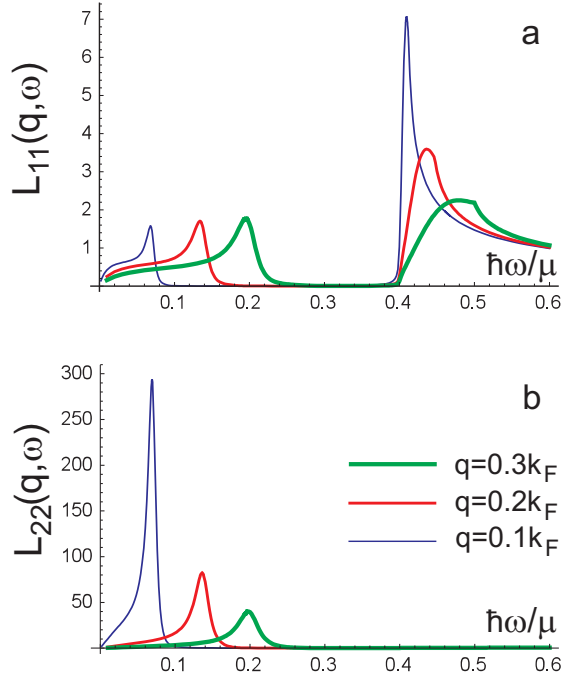}
\end{center}
\caption{Frequency dependence of the energy loss functions (\ref{58-1}), (\ref{58-2}) in
$\mu/gk_F^2$ units at $T=0.1\mu$, $q=0.1 k_F,\ 0.2 k_F,\ 0.3 k_F$, $\Delta=0.2\mu$ and
$\gamma=10^{-3}\mu$} \label{f3}
\end{figure}

\begin{figure}
\begin{center}
\includegraphics[width=8cm]{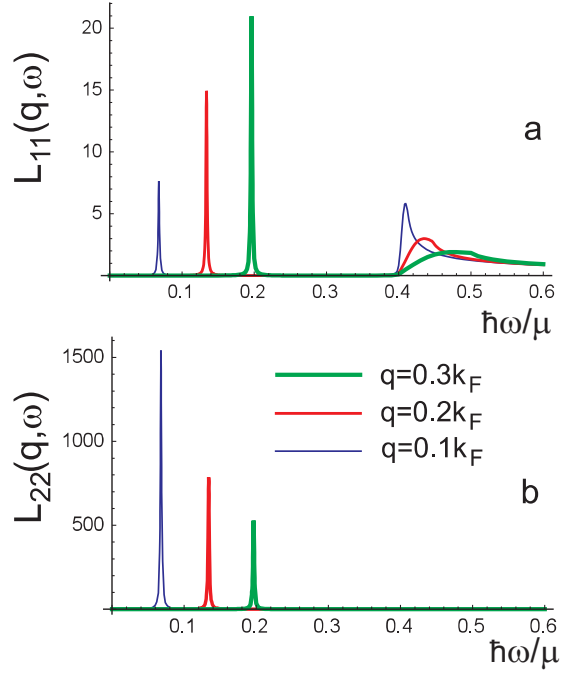}
\end{center}
\caption{The same as in Fig. \ref{f3} at $T=0$.} \label{f4}
\end{figure}

The frequency dependencies of $L_{11}(\mathbf{q},\omega)$ and $L_{22}(\mathbf{q},\omega)$
at three different $q$ and $T=0.1\mu$ are shown in Fig. \ref{f3}. One can see that the
function $L_{22}(\mathbf{q},\omega)$, Fig. \ref{f3}(b), has a peak at $\hbar\omega<
2\Delta$. The function $L_{11}(\mathbf{q},\omega)$, Fig. \ref{f3}(a), has two peaks, one
is at $\hbar\omega< 2\Delta$ (at the same frequency as the peak in Fig. \ref{f3}(b)) and
the other, at $\hbar\omega> 2 \Delta$. Two peaks in Fig. \ref{f3}a appear due to the
coupling of oscillations of the amplitude and the phase of the order parameter (in
conventional superconductors these oscillations are decoupled from each other \cite{35}).
In Fig. \ref{f4} we present the same  dependencies as in Fig. \ref{f3} at $T=0$. One can
see that the positions of the peaks remain practically unchanged under lowering of
temperature (at $\Delta=const$). At the same time an essential narrowing of the
low-frequency peak at $T=0$ signals a decrease of the damping rate of the lower mode. It
is connected with the fact that the Landau damping in the frequency domain $\hbar\omega<
2 \Delta$ is proportional to $\exp(-\Delta/T)$. In contrast, in the frequency domain
$\hbar\omega> 2 \Delta$ the Landau damping remains strong even at $T=0$. Therefore the
high-frequency peak is not changed under lowering of temperature.

\begin{figure}
\begin{center}
\includegraphics[width=8cm]{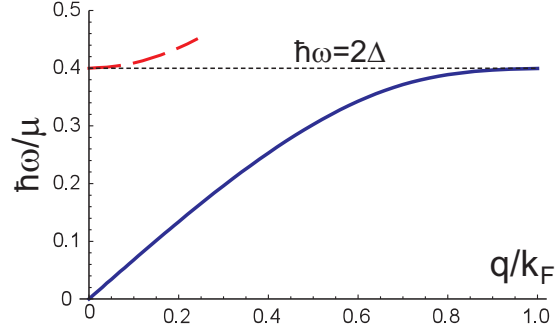}
\end{center}
\caption{The spectra of the Anderson-Bogoliubov (solid curve) and Schmid (dashed curve)
modes in the double layer graphene system.} \label{f4-1}
\end{figure}

The spectra of the modes determined by Eq. (\ref{57}) are shown in Fig. \ref{f4-1}. The
dependencies presented are obtained from the position of the maximum of the functions
(\ref{58-1}) and (\ref{58-2}) at $T=0$. At small wave vectors the dispersion relation for
the lower mode is approximated by the expression $\omega= q v_F/\sqrt{2}$, that is the
spectrum of the AB mode in two-dimensions. At large $q$ the frequency of this mode
approaches $\omega=2\Delta/\hbar$. The frequency of the upper mode approaches
$2\Delta/\hbar$ at $q\to 0$. This mode can be recognized only in the limit $q/k_F\ll 1$.
At $q/k_F\gtrsim 0.2$ the peak that corresponds to that mode washes out. The lower mode
in  Fig. \ref{f4-1} should be interpreted as an analog of the AB mode and the upper mode,
as the analog of the Schmid mode.

The third  multiplier in the determinant of the matrix in Eq. (\ref{49}) yields the
equation
\begin{eqnarray}\label{58}
\left[\Pi_{11}^{(R)}(\mathbf{q},\omega)
\Pi_{22}^{(R)}(\mathbf{q},\omega)+[\Pi_{12}(\mathbf{q},\omega)]^2\right]
\left[1-V_-(q)\Pi_{33}(\mathbf{q},\omega)\right] \cr
-V_-(q)\left[\Pi_{11}^{(R)}(\mathbf{q},\omega)[\Pi_{23}(\mathbf{q},\omega)]^2
-\Pi_{22}^{(R)}(\mathbf{q},\omega)[\Pi_{13}(\mathbf{q},\omega)]^2+
2\Pi_{12}(\mathbf{q},\omega)\Pi_{13}(\mathbf{q},\omega)\Pi_{23}(\mathbf{q},\omega)\right]=0.
\end{eqnarray}
One can see that at $V_-(q)=0$ (that corresponds to $d=0$)  Eq. (\ref{58}) coincides with
Eq. (\ref{57}).

At $V_-(q)\ne 0$ Eq. (\ref{58}) can be rewritten in the form
\begin{equation}\label{50-1}
    \varepsilon_-(\mathbf{q},\omega)=1-V_-(q)\Pi_{-}(\mathbf{q},\omega)=0,
\end{equation}
where
\begin{eqnarray}\label{82}
\Pi_{-}(\mathbf{q},\omega)=
\Pi_{33}(\mathbf{q},\omega)+\frac{\Pi_{11}^{(R)}(\mathbf{q},\omega)
[\Pi_{23}(\mathbf{q},\omega)]^2
-\Pi_{22}^{(R)}(\mathbf{q},\omega)[\Pi_{13}(\mathbf{q},\omega)]^2+
2\Pi_{12}(\mathbf{q},\omega)\Pi_{13}(\mathbf{q},\omega)\Pi_{23}(\mathbf{q},\omega)}
{\Pi_{11}^{(R)}(\mathbf{q},\omega)\Pi_{22}^{(R)}(\mathbf{q},\omega)+[\Pi_{12}(\mathbf{q},\omega)]^2}.
\end{eqnarray}
 The function $\Pi_{-}(\mathbf{q},\omega)$ can be
understood as the polarization function "dressed"\ by the order parameter oscillations.
 {Numerical evaluation confirms the gauge invariance of the function
(\ref{82}) in the limit $\gamma\to 0$ (see the Appendix).}

Equation (\ref{50-1}) is the dispersion equation for antisymmetric plasma oscillations
coupled to the order parameter oscillations. The dielectric function
$\varepsilon_-(\mathbf{q},\omega)$ determines screening of the scalar potential of a test
charge $\rho^{\mathrm{test}}_-$: $e^2\varphi^{\mathrm{scr}}_-(\mathbf{q},\omega)=V_{-}(q)
\rho^{\mathrm{test}}_-(\mathbf{q},\omega)/{\varepsilon_-(\mathbf{q},\omega)}$.

The relation between the polarization function $\Pi_{-}(\mathbf{q},\omega)$ and the
counterflow conductivity is given by the equation
\begin{equation}\label{55-1}
    \sigma_{-,xx}(q \mathbf{i}_x,\omega)
=\frac{ie^2\omega}{q^2}\Pi_{-}(q \mathbf{i}_x,\omega).
\end{equation}
Using the condition of the gauge invariance (\ref{a3}) and the expressions (\ref{a7}) and
 (\ref{a5}) in the Appendix,  one can show that at small $q$ the quantity $\Pi_{-}(q
\mathbf{i}_x,\omega)\propto q^2$. Therefore the conductivity $\sigma_{-,xx}(q
\mathbf{i}_x,\omega)$  given by Eq. (\ref{55-1}) is finite at $q \to 0$ .  Also from the
physical reasons the real part of $\sigma_{-,xx}(q \mathbf{i}_x,\omega)$ should be
positive. We have checked numerically the fulfillment of the latter condition.

The dispersion equation for the antisymmetric (acoustic) plasmon mode that accounts for
retarded effects  has the form \cite{31,36}
\begin{equation}\label{72}
    \left(1+\frac{4\pi i \kappa_1
\sigma_{-,xx}(q\mathbf{i}_x,\omega)}{\omega}\right)\tanh \frac{\kappa_2 d}{2}
+\frac{\varepsilon \kappa_1}{\kappa_2}=0.
\end{equation}
In the nonretarded approximation ($\kappa_1=\kappa_2=q$) Eq.
 (\ref{72}) reduces
to Eq. (\ref{50-1}).

\begin{figure}
\begin{center}
\includegraphics[width=8cm]{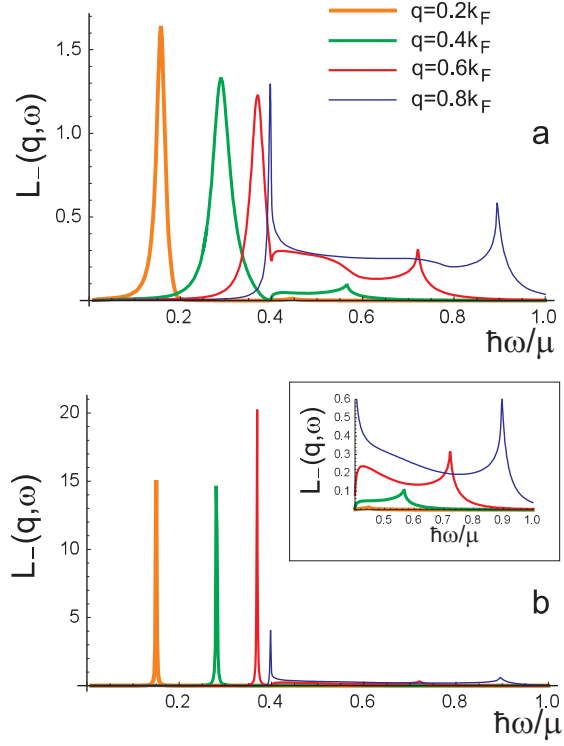}
\end{center}
\caption{Frequency dependence of the energy loss function (\ref{52-1}) at $T=0.1\mu$ (a)
and $T=0$ (b). The high-frequency peaks at $T=0$ are shown in the inset in another scale.}
\label{f5}
\end{figure}

We analyze Eq. (\ref{50-1}) considering the energy loss function
 \begin{equation}\label{52-1}
    L_-(\mathbf{q},\omega)=-\mathrm{Im}\left[\frac{1}{\varepsilon_-(\mathbf{q},\omega)}\right].
 \end{equation}
The frequency dependencies of  $L_-(\mathbf{q},\omega)$ at four different wave vectors
($q=0.2 k_F,\ 0.4 k_F,\ 0.6 k_F,\ 0.8 k_F)$,  $\Delta=0.2\mu$, $T=0.1 \mu$ and $T=0$ are
shown in Fig. \ref{f5}. The parameters used for the calculations are $\varepsilon=4$, $d
k_F=0.1$ and $\gamma=10^{-3}\mu$. One can see that in similarity with
$L_+(\mathbf{q},\omega)$ the function $L_-(\mathbf{q},\omega)$ contains two peaks, one is
below the gap $2\Delta$ and the other, above the gap. The low-frequency peak is narrower
than the high-frequency one. Differently from the $L_+(\mathbf{q},\omega)$-dependence the
position of the lower peak of the $L_-(\mathbf{q},\omega)$-dependence remains practically
unchanged under variation of temperature (at $\Delta= const$). This peak does not
disappear at $T=0$.

In conventional superconductors the mode  that corresponds to coupled oscillations of the
scalar potential and the phase of the order parameter is known as the Carlson-Goldman
(CG) mode\cite{40}. The frequency of the CG mode satisfies the inequality
$\hbar\omega<2\Delta$. The mentioned similarities allow us to interpret  the lower
antisymmetric mode as an analog of the Carlson-Goldman mode.

Lowering of temperature results in a considerable decrease of the damping rate of the
lower mode but does not influence the damping rate of the upper mode. As in the case of
the AB and Schmid modes, it is connected with the specific temperature and frequency
dependence of the Landau damping in the state with electron-hole pairing\cite{31}.

\begin{figure}
\begin{center}
\includegraphics[width=8cm]{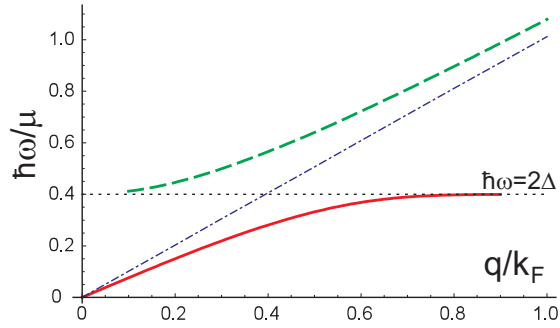}
\end{center}
\caption{The dispersion curves for the  Carlson-Goldman mode (solid curve) and for the
upper
 antisymmetric plasmon mode (dashed curve). The spectrum of the antisymmetric (acoustic)
 plasmon mode in the normal state is
shown by  the dash-dotted line.} \label{f5-1}
\end{figure}

 The
dispersion curves  calculated from the positions of two maxima of the function
(\ref{52-1})  at $T=0$ are shown in Fig. \ref{f5-1}. The lower mode has the acoustic
dispersion relation at small wave vectors. At large $q$ its frequency approaches
$2\Delta/\hbar$. The dispersion curve for the acoustic plasmon mode in the normal state
($\Delta=0$) calculated at  the same parameters is also shown in Fig. \ref{f5-1}. It is
known\cite{pl} that the velocity $v_a$ of the acoustic plasmon in a double-layer graphene
system can be very close to $v_F$, but it is always larger than $v_F$ irrespective of the
values of $d$ and $\varepsilon$. For $\varepsilon$ and $d$ specified above $v_a\approx
1.016 v_F$. The velocity of the CG mode $v_{CG}$ can be smaller than $v_F$. In our case
$v_{CG}\approx 0.77 v_F$. The velocity  $v_{CG}$ is larger than the velocity of the AB
mode $v_{AB}=v_F/\sqrt{2}$ but there is no requirement for $v_{CG}$ to be larger than
$v_F$. It is correlated with the fact that in the normal state the mode with the phase
velocity $v_{ph}<v_F$ should experience strong Landau damping, but in the paired state
the modes with $\hbar\omega<2 \Delta$ do not experience Landau damping at $T=0$.

At  $q/k_F>0.9$ the peak at the $L_-(\mathbf{q},\omega)$-dependence that corresponds to
the CG mode disappears. In contrast, the upper mode peak is well recognized at large $q$,
while at small $q$ this peak almost disappears. At $\Delta\to 0$ the upper mode is
transformed into the acoustic plasmon mode. It allows us to interpret the upper
antisymmetric mode  as a residual acoustic plasmon mode.

 {It is instructive to compare the CG mode in conventional superconductors
and in counterflow superconductors. Under the two-fluid picture, the CG mode is regarded
as out-of-phase motion of the superfluid and normal components. In conventional $s$-wave
superconductors the CG mode can be observed only at the temperature close to the critical
temperature $T_c$\cite{35, cg97}. At such temperatures the density of the normal
component is comparable to the density of the superfluid component. But in clean $s$-wave
superconductors at $T$ close to $T_c$  the CG mode is smeared out due to the Landau
damping of the quasiparticles and it can be clearly seen only in dirty
systems\cite{cg97}. In $d$-wave superconductors, due to the presence of four Fermi points
at the nodes of the $d$-wave order parameter the CG mode can be registered in clean
systems and it survives at much lower temperatures, down to $T\sim 0.1 T_c$
\cite{cg00,cg02}. The CG mode was also predicted for a color–flavor locked (CFL) phase of
color superconducting dense quark matter \cite{cg02-1}. The presence of two different
types of quarks with nonequal gaps in the CFL phase causes a partial suppression of the
Landau damping. As a consequence, the CG mode can be observed in the pure limit at a
temperature close  to the critical one ($T/T_c\geq 0.986$)\cite{cg02-1}. The situation in
counterflow superconductors differs from ones in $s$-wave and $d$-wave superconductors
and for a CFL phase of superconducting quark matter. In the counterflow superconductors
the CG mode can be interpreted as in-phase motion of the superfluid and normal
components, and due to that the CG mode can be observed at all temperatures below the
critical one, in particular  at $T=0$ (Fig. \ref{f5-1}). At low temperature the Landau
damping is suppressed. Therefore we consider pure counterflow superconductors as more
appropriate for the observation of the CG mode.}

 { In this study we consider the contact pairing potential. For more
careful analysis the contact potential should be replaced with a screened Coulomb
potential. In this case one should take into account a dependence of the order parameter
on the momentum (see Sec. \ref{s2}). To describe the state with the momentum-dependent
order parameter, one can approximate the screened Coulomb interaction by a function which
is separable in the incoming and outgoing momenta, as was done in Refs. \onlinecite{17,
ls09}. Restricting with the separable pairing potential and considering the close-band
pairing (the pairing of carriers in the conduction band in layer 1 with carriers in the
valence band in layer 2) we arrive at the polarization functions with the additional
momentum-dependent factor under the integral over $\mathbf{k}$. Similar factor emerges in
the polarization functions for $d$-wave superconductors \cite{cg00} (in the latter case
this factor is angle-dependent). Evaluating  the polarization functions with the
additional factor, we obtain  dispersion curves that are very close to ones obtained for
the model with the contact pairing potential. It can be understood as follows. The
collective mode spectra presented in Figs. \ref{f4-1} and \ref{f5-1} are determined in
the main part by two parameters: the Fermi velocity $v_F$ and the gap in the
quasiparticle spectrum $2\Delta$. The parameter $2\Delta$ is sensitive to the form of the
pairing potential, and it depends on the interlayer distance and on the density of the
carriers. But in our study we do not evaluate this parameter. We just fix its value. If
the parameter $2\Delta$ is fixed,  accounting for a momentum dependence of the order
parameter does not influence significantly the collective mode spectrum. Thus we conclude
that the model with the contact pairing potential adequately describes collective modes
in counterflow superconductors.}

\section{Conclusion}

In conclusion, we have shown that explicit accounting for the order parameter
oscillations is crucial in obtaining the spectra of antisymmetric plasma modes in  double
layer systems with electron-hole pairing. At the same time the approach \cite{31} based
on a particular solution of the generalized Ward identity cannot describe a number of
important features. In particular, taking into account the order parameter oscillations,
we predict the existence of two antisymmetric modes. The upper mode can be interpreted as
a residual normal state acoustic plasmon, and the lower mode, as an analog of the
Carlson-Goldman mode. Two more modes interpreted as analogs of the Anderson-Bogoliubov
and Schmid modes are also identified. The latter modes are associated with out-of-phase
oscillations of the order parameters of  two spin subsystems.

While the results are obtained with reference to a double monolayer graphene, one can
expect that they reflect the general collective mode behavior in double layer systems
with electron-hole pairing. Our approach can be easily extended to the double bilayer
graphene systems\cite{20,21}. The polarization functions for the double bilayer graphene
are obtained from Eq. (\ref{39}) under substitutions $\xi_{\mathbf{k},\lambda} \approx
\lambda\hbar^2 k'^2/2m-\mu$ and $\chi_\mathbf{k} \approx \mp 2\theta_{\mathbf{k}'}$,
where $m$ is the effective mass. Preliminary calculations show that the collective mode
systematics for the double bilayer graphene systems is the same as for the double
monolayer ones. At the same time we emphasize that our approach is not applied to the
systems with low density of carriers and a large gap between the valence and conduction
bands.  The counterflow superconductivity  in the low density limit is described by the
interacting boson model\cite{24,41,42}. Such systems also have two superfluid components
but the frequency of the mode that corresponds to out of phase oscillations of two
components becomes imaginary-valued under increase of the interlayer distance
\cite{24,42}.  It signals an instability with respect to spatial separation of the
components. The system considered in the present paper does not show softening of the
out-of-phase mode and it is stable with respect to spatial separation.

\section*{Acknowledgment}
This study was supported by a grant of the Ukraine State Fund for Fundamental Research
(Project No. 33683).

\appendix
\label{ape}
\section{Gauge invariance of the polarization functions}
 { Assuming that  the $x$-components of the vector potential
$\mathbf{A}_{\pm}=\mathbf{A}_1\pm\mathbf{A}_2$ are nonzero one can obtain the following
expression for the charge density oscillations
\begin{equation}\label{a1}
    \rho_{\pm}(q \mathbf{i}_x,\omega)=e^2 \left[\Pi_{\pm,0}(q \mathbf{i}_x,\omega)
    \varphi_{\pm}(q \mathbf{i}_x,\omega)+\Pi_{\pm,x}(q \mathbf{i}_x,\omega)
    A_{\pm,x}(q \mathbf{i}_x,\omega)\right].
\end{equation}
Here $\Pi_{+,0}( \mathbf{q},\omega)\equiv \Pi_{00}( \mathbf{q},\omega)$ and $\Pi_{-,0}(
\mathbf{q},\omega)\equiv \Pi_{-}( \mathbf{q},\omega)$ are the polarization functions
given by Eqs. (\ref{39}) and (\ref{82}). The functions $\Pi_{\pm,x}(q
\mathbf{i}_x,\omega)$  in Eq. (\ref{a1}) describe the response to the vector potential
(the interaction with the vector potential is given by the Hamiltonian $H_{A}=-(1/2c)\int
d\mathbf{r} [j_{+,x}A_{+,x}+j_{-,x}A_{-,x}]$). The explicit expressions for these
quantities are the following:
\begin{equation}\label{a2}
\Pi_{+,x}(\mathbf{q},\omega)=\frac{1}{S}\sum_{\nu_1,\nu_2}
\delta_{\mathbf{k}_1-\mathbf{q},\mathbf{k}_2} \Phi^{0 3}_{\nu_1\nu_2}
\frac{\lambda_1\cos\chi_{\mathbf{k}_1}+\lambda_2\cos\chi_{\mathbf{k}_2}}{2}
\frac{f_{\nu_1}-f_{\nu_2}}
    {E_{\nu_1}-E_{\nu_2}-\hbar(\omega+i\gamma)}
    \end{equation}
and
\begin{eqnarray}\label{a7}
\Pi_{-,x}(\mathbf{q},\omega)=\frac{v_F}{c}\left[\Pi_{33,x}-\frac{\Pi_{31,x}
[\Pi_{13}\Pi^{(R)}_{22}-\Pi_{12}\Pi_{23}]
+\Pi_{32,x}[\Pi_{23}\Pi_{11}^{(R)}-\Pi_{21}\Pi_{13}]} {\Pi_{11}^{(R)}\Pi_{22}^{(R)}-
\Pi_{12}\Pi_{21}}\right],
\end{eqnarray}
where the functions $\Pi_{ss'}(\mathbf{q},\omega)$ and
$\Pi_{ss}^{(R)}(\mathbf{q},\omega)$ are given by Eqs. (\ref{39}), (\ref{43}), and
\begin{equation}\label{a5}
\Pi_{3s,x}(\mathbf{q},\omega)=\frac{1}{S}\sum_{\nu_1,\nu_2}
\delta_{\mathbf{k}_1-\mathbf{q},\mathbf{k}_2} \Phi^{s 0}_{\nu_1\nu_2}
\frac{\lambda_1\cos\chi_{\mathbf{k}_1}+\lambda_2\cos\chi_{\mathbf{k}_2}}{2}
\frac{f_{\nu_1}-f_{\nu_2}}
    {E_{\nu_1}-E_{\nu_2}-\hbar(\omega+i\gamma)}
    \end{equation}
($s=1,2,3$).}

 { The gauge invariance  requires that
\begin{equation}\label{a3}
    \omega \Pi_{\pm,0}(q \mathbf{i}_x,\omega)-qv_F\Pi_{\pm,x}(q \mathbf{i}_x,\omega)=0.
    \end{equation}
Numerical evaluation of the left-hand part of (\ref{a3}) with the upper as well as with
the lower sign shows that  it goes to zero at $\gamma\to 0$. Thus we conclude that in the
pure limit our approach yields the gauge invariant polarization functions.}

\end{document}